\documentclass[aps,prl,twocolumn,floats,showpacs,superscriptaddress]{revtex4}
\usepackage{graphicx,epsfig}% Include figure files
\usepackage{times}
\begin{document}
\unitlength 1 cm
\newcommand{\be}{\begin{equation}}
\newcommand{\ee}{\end{equation}}
\newcommand{\bearr}{\begin{eqnarray}}
\newcommand{\eearr}{\end{eqnarray}}
\newcommand{\nn}{\nonumber}
\newcommand{\vk}{\vec k}
\newcommand{\vp}{\vec p}
\newcommand{\vq}{\vec q}
\newcommand{\vkp}{\vec {k'}}
\newcommand{\vpp}{\vec {p'}}
\newcommand{\vqp}{\vec {q'}}
\newcommand{\bk}{{\mathbf k}}
\newcommand{\bp}{{\mathbf p}}
\newcommand{\bq}{{\mathbf q}}
\newcommand{\br}{{\mathbf r}}
\newcommand{\bR}{{\mathbf R}}
\newcommand{\bd}{{\mathbf d}}
\newcommand{\up}{\uparrow}
\newcommand{\down}{\downarrow}
\newcommand{\fns}{\footnotesize}
\newcommand{\ns}{\normalsize}
\newcommand{\cdag}{c^{\dagger}}

\title{Fermi surface nesting and possibility of orbital ordering in FeO}

\author{M. Alaei}
\affiliation{Department of Physics, Isfahan University of
Technology, Isfahan 84154-83111, Iran}
%\author{S. A. Jafari{\footnote {Electronic address:
%sa.jafari@cc.iut.ac.ir}}}

\author{S. A. Jafari}
\affiliation{Department of Physics, Isfahan University of
Technology, Isfahan 84154-83111, Iran}
\affiliation{School of Physics, Institute for Research in Fundamental Sciences (IPM), Tehran 19395-5531, Iran}

\begin{abstract}
We study FeO, a Mott insulator in GGA and GGA+U approximations.
In the GGA we find a multi-band metallic state with remarkable inter-band nesting between 
two $t_{2g}$ bands of Fermi surface, which signals possible instability
towards an orbital ordered insulating phase.
Such broken symmetry state, although has lower energy than the underlying 
homogeneous metallic state, but the gap magnitude is 
less than the experimentally observed optical gap. Therefore we incorporate the
calculated value of on-site Coulomb repulsion U on orbital ordered state.
We find that symmetry breaking and Coulomb correlations cooperate together
to stabilize the system and give an insulating orbital ordered state,
with the gap magnitude very close to the experimental value.
We propose this method as a possible indication of orbital ordering in LDA and
GGA calculations. We check our method with known examples of LiVO$_2$ and LaMnO$_3$.
\end{abstract}

\maketitle

\section{introduction}
Charge and spin are two fundamental characteristics of electrons. 
Pauling invented the concept of "orbital" as another important attribute 
of electrons. This theoretical tool
which intuitively refers to the shape of electronic clouds,
was so powerful that enabled a qualitative understanding of the 
electronic properties  
of elements in the periodic table, as well as binding in many 
molecules and solids in very early days of quantum 
mechanics, when no computers existed. 
Unlike charge and spin which directly couple to many 
experimental probes, the question of the experimental 
observation of orbital degrees of freedom remained 
untouched until recently~\cite{IshiharaReview}.
Resonance x-ray scattering has been successfully employed for
direct observation of the ordering of orbitals in manganites, 
A$_{1-x}$B$_x$MnO$_3$~\cite{Murukami98}, as well as some other
systems such as transition metal oxides 
(See table 1 of Ref.~\onlinecite{IshiharaReview}).

   Observation of orbital ordering has established that
orbitals are not just theoretical hypotheses.
They are another degree of freedom, just like charge and
spin which at low temperatures can form various types of ordered states. 
For materials such as KCuF$_3$ and LaMnO$_3$, 
which are {\em a priori} known to be orbital ordered, 
people have employed LDA+U
approximation with the assumption of orbital ordering to
justify the ground state properties~\cite{Anisimov95,Korotin}.
However, an important
theoretical question is, given the Fermi surface topology
and the electronic structure of an unknown material, under
what conditions the strong correlation effects are anticipated
to stabilize an orbital ordered state at low temperatures.
In this paper we show that one can 
study the possibility of orbital ordering within {\em ab initio}
spin-density-functional scheme by appropriate Fermi surface
analysis. We employ our method to predict the possibility of
orbital ordered phase for FeO at ambient pressure. 

Despite the stoichiometric simplicity, the iron-monoxide, FeO,
presents a challenge in terms of the theoretical understanding
of its electronic and magnetic properties~\cite{Mazin,Blaha,Gironcoli}.
  In series of transition metal mono-oxides the cubic crystal field
splits the $d$ orbitals into two-fold $e_g$ and three-fold
degenerate $t_{2g}$  states. As one moves from MnO to NiO along
the periodic table of elements, the minority spins are being
filled. In these systems, each (111) plane has majority spins 
aligned ferromagnetically within the plane
(and so are the minority spins),
while in the next
(111) plane the ferromagnetic alignment is in opposite direction.
When we consider the filling of minority spins, the degeneracy
left behind in $t_{2g}$ gives rise to a metallic state with
partially filled $t_{2g}$ band for FeO and CoO in Density
functional Theory (DFT) within generalized gradient approximation
(GGA). However, this prediction is in contradiction to
experiment, where it is found to be a Mott insulator with an 
optical band gap of $2.4$ eV~\cite{Bowen}.
Ref.~\onlinecite{Alfredsson} summarizes the various theoretical
values obtained for the optical gap. The Hartree-Fock approximation
largely overestimates the gap~\cite{Alfredsson}. The diffusion
Monte Carlo method give the gap value of $2.8\pm 0.3$ eV~\cite{MitasFeO}.
In this paper, we show that this insulating behavior
can be accounted for by a orbital ordering, on top of which
a Hubbard type of correlation further stabilizes the 
insulating state. The gap value we obtain is $2.2$ eV.
In this  research we use QUANTUM-ESPRESSO~\cite{qe} as DFT code to apply
GGA and GGA+U to FeO and GGA to LiVO$_2$ and LaMnO$_3$.
For core-electron
interaction ultra-soft pseudopotentials have been used.
Wavefunction and charge density are expanded in plane-wave with $40$ Ry
and $400$ Ry cutoff respectively. The Hubbard U is set to $4.3$ eV for
GGA+U calculation for FeO~\cite{Gironcoli}.
%%%%%%%FIG1
\begin{figure*}[t]
   %\vspace{0.7 cm}
    \includegraphics[width=14cm,angle=0]{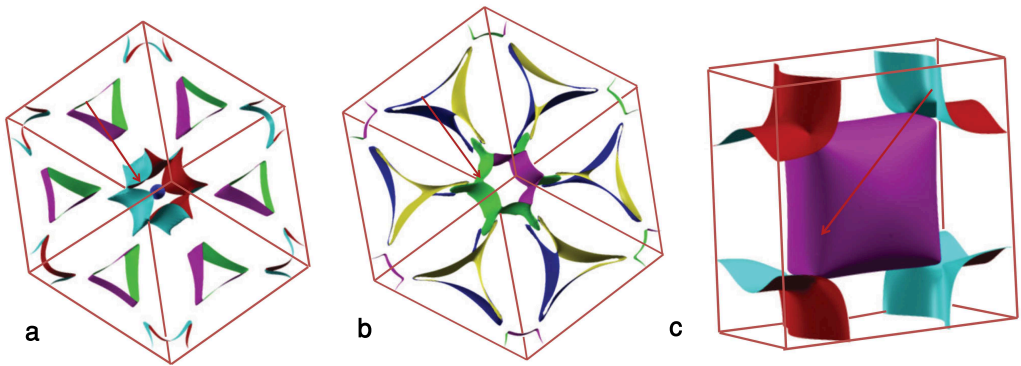}
    \caption{(Color online) Fermi surface for (a) FeO, (b) $\rm LiVO_{2}$ and (c) 
    $\rm LaMnO_{3}$. The Fermi surfaces are plotted in the reciprocal lattice (not 
    in the first Brillouin zone). For FeO and $\rm LiVO_{2}$ the lattice is 
    trigonal and for $\rm LaMnO_{3}$ is tetragonal. 
    %For FeO and $\rm LaMnO_{3}$ 
    %we used a antiferromagnetic phase so the Fermi surfaces of them belongs to 
    %spin up (the Fermi surface of spin down are the same). 
    %For $\rm LiVO_{2}$ we used non-magnetic phase 
    }
    \label{fig:all_fs}
\end{figure*}
%%%%%%%FIG1

Nested Fermi surfaces can potentially lower the symmetry
of the ground state by either developing some sort of long-range order~\cite{Gruner},
or appropriate distortion of the lattice structure~\cite{Spaldin}. 
Efremov and Khomskii~\cite{Khomskii2005}  developed a theory of 
orbital ordering in terms of inter-band nesting between the
two $e_g$ bands in manganites. According to their work, the 
underlying inter-band nesting can lead to instability towards
a symmetry broken phase with ordering in $e_g$ orbitals.
In this work we analyze the GGA Fermi surface of FeO. 
 We observe
a remarkable nesting pattern between {\em two of} the three ($t_{2g}$)
bands crossing the Fermi level. 
This qualitative observation can
be made quantitative by calculating appropriate inter-band
susceptibility which peaks around ${\mathbf q}=1/3{\mathbf
G}$, where $\mathbf G$ is the reciprocal lattice vector in (111)
plane. The evidence for instability towards orbital ordered phase from GGA
metallic ground state of FeO is substantiated with another
independent GGA calculation for the broken symmetry phase. 
We also perform a GGA+U calculation to compare
states with broken symmetry and those without broken symmetry.
Thus, orbital ordering (OO), and Hubbard correlations work together to give 
a Mott insulator with orbital ordering pattern, in agreement
with a similar scenario proposed for LiVO$_2$~\cite{Pen}.

%In this  research we used QUANTUM-ESPRESSO as DFT code to apply
%GGA and GGA+U to FeO. 
%We apply GGA and  GGA+U to study FeO.
To verify our approach, we apply this analysis to the well known 
orbital ordered materials, such as,
$\rm LiVO_{2}$ and $\rm LaMnO_{3}$. The former example is
closer to FeO in two respects: (i) The ordering pattern occurs
in a triangular lattice of $(111)$ plane. (ii) Both LiVO$_2$
and FeO have degenerate $t_{2g}$ bands; however, the difference
is that in LiVO$_2$ we have a $d^2$ class~\cite{Pen}, while 
the minority spin electrons of FeO belong to $d^1$ class.
Note that the majority spin states are well separated from those of
minority spins~\cite{Gironcoli}.
%Orbital ordering in $d^2$ triangular lattice underlying $t_{2g}$ bands
%of LiVO$_2$ has been considered by Pen and coworkers~\cite{Pen}. They find
%two mean field states (Fig. 1 of Ref.~\onlinecite{Pen}). 
In case of FeO we are
dealing with Fe$^{2+}$ cations with $3d^6$ configuration. By Hund's rule,
such $d^6$ configuration is equivalent to $d^1$ class in the minority spin
sector. 
%According to Goodenough-Kanamori-Anderson rules, the 
%less than half-filled configuration of $d^1$ class implies ferromagnetic 
%spin coupling in the $(111)$ plane of Fe ions.
We first calculate the GGA bands and Fermi surface for the trigonal structure.
This gives us a metallic state with remarkable inter-band nesting 
shown in Fig.~\ref{fig:all_fs}(a). In this figure, the nesting 
is only between two of the $t_{2g}$ Fermi surfaces. The third 
$t_{2g}$ band does not contribute substantially to the Fermi
surface. We have checked that this band is further pushed
away from the Fermi level by small amounts of on-site Coulomb
interaction $U$. Hence despite the threefold degeneracy 
of $t_{2g}$ bands, here we are dealing with a two-band situation
to which the picture proposed by Efremov and Khomskii in the
context of orbital ordering instability of $e_{g}$ bands~\cite{Khomskii2005} 
can be applied.
The same type of inter-band nesting can be observed in the
GGA Fermi surfaces of LiVO$_2$ (Fig.~\ref{fig:all_fs}(b)),
which is known to exhibit orbital ordering at low temperatures~\cite{Pen,Yu,PenPRB}.
The example of LiVO$_2$ is very similar to FeO, in terms of underlying triangular
lattice structure  in $(111)$ plane. This is also reflected in 
their Fermi surface topologies in Fig.~\ref{fig:all_fs}.
In Fig.~\ref{fig:all_fs}(c) we plot the GGA Fermi surface of 
LaMnO$_3$. As shown by Efremov and Khomskii~\cite{Khomskii2005},
such a nesting leads to an instability towards orbital ordered
state at low enough temperatures.
In the following we are going to argue that the same argument
predicts orbital ordering for LiVO$_2$ and FeO.

To quantify the above discussion, we calculate the
contribution of inter-band processes to generalized susceptibility
defined by,
\begin{equation}
   \chi_{\rm inter}(\mathbf q)=\sum_{\mathbf{k}} \frac{f(\epsilon^a_\mathbf{k})[1-f(\epsilon^b_\mathbf{k+q})]}
   {\epsilon^b_\mathbf{k+q}-\epsilon^a_{\mathbf{k}}},
   \label{GS.eqn}
\end{equation}
where $a$ and $b$ denotes the two nested bands, 
$f(\epsilon)$ is the occupation number and $\epsilon^{a,b}_\mathbf{k}$ 
denotes the band energy. To identify 
the exact value of nesting vector we calculated generalized
susceptibility in direction of $\mathbf{G_{1}\pm G_{2}}$, where
$\mathbf G_{1}$ and $\mathbf G_{2}$ are two basis vectors in the 
reciprocal space. 
%%%%%%%FIG1
\begin{figure}[h]
    \includegraphics[width=8cm,angle=0]{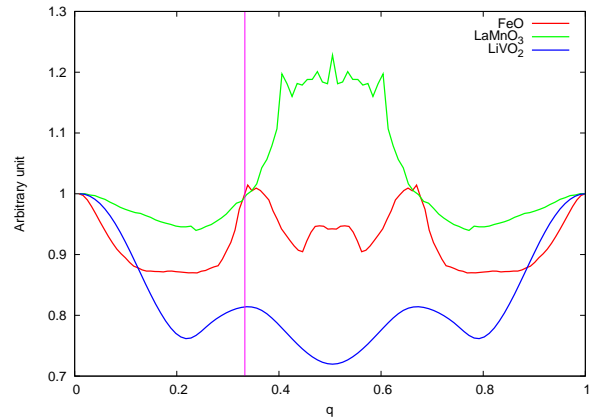}
    \caption{(Color online) Generalized susceptibility for FeO, $\rm LiVO_{2}$ 
    and $\rm LaMnO_{3}$. For FeO and $\rm LiVO_{2}$ the direction of q is parallel 
    to $\mathbf{G}_1-\mathbf{G}_2$ where $\mathbf{G}_1$ and $\mathbf{G}_2$ are 
    vectors in reciprocal space of the trigonal lattice. For $\rm LaMnO_{3}$ we chose q in the direction 
    of $\mathbf{G}_1+\mathbf{G}_2$ where $\mathbf{G}_1$ and $\mathbf{G}_2$ are 
    reciprocal lattice vectors of the tetragonal lattice ($\mathbf{\mid G }_1\mid=\mathbf{\mid G}_2\mid$). 
    The vertical line indicates the maximum of generalized susceptibility 
    at $\mathbf{q}=\frac{1}{3}(\mathbf{G}_1-\mathbf{G}_2)$ for  FeO and $\rm LiVO_{2}$.}
    \label{fig:GS}
\end{figure}
%%%%%%%FIG1

In Fig.~\ref{fig:GS} we plot the results of numerical evaluation of
integral (\ref{GS.eqn})  for values of $\mathbf q$ in the direction 
indicated by arrows on panels (a)-(c) of Fig.~\ref{fig:all_fs}, respectively~\cite{tech-details}.
As can be seen in Fig.~\ref{fig:GS}, for the case of LaMnO$_3$, the 
inter-band susceptibility peaks around the ordering vector, 
$\mathbf{q}=\frac{1}{2}(\mathbf{G}_1+\mathbf{G}_2)$
in $(001)$ plane~\cite{Khomskii2005}.
This implies an ordering in the $xy$ plane with $2a\times 2a$ unit cell.
Now let us focus on the case of $t_{2g}$ systems, FeO and LiVO$_2$.
As can be seen in Fig.~\ref{fig:GS}, for the case of FeO and LiVO$_2$, 
the susceptibility peaks around 
$\mathbf{q}=\frac{1}{3}(\mathbf{G}_1-\mathbf{G}_2)$.
Based on the theory of Ref.~\cite{Khomskii2005}, we speculate 
this implies a possible instability towards ordering of $t_{2g}$ bands 
on a triangular lattice of $(111)$ plane.

Feeding these information back into the DFT machinery, we 
examine the total energy, gap magnitude and charge distribution
profile assuming an ordering pattern suggested from
our Fermi surface analysis.
Hence we are lead to examine a broken symmetry phase with a 
$\sqrt 3 a \times \sqrt 3 a$ supercell in $(111)$ 
plane~\cite{Pen,Gironcoli}.
The structure contains three Fe atom per (111) plane. 
%%%%%%%FIG1
\begin{figure}[t]
    \includegraphics[width=8cm,angle=0]{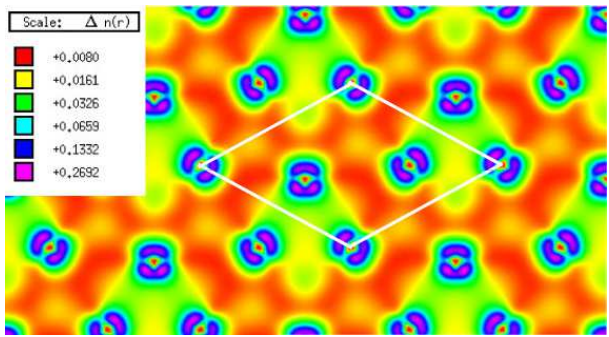}
    \caption{(Color online) Charge density of minority spin of Fe atoms in 
    (111) surface with GGA calculation. The iron atoms are slightly displaced 
    and then they are allowed to be relaxed. As can be seen, the new minimum 
    in this GGA calculation corresponds to an orbital ordered state.
    The new cell  shown in this figure is $\sqrt 3 a\times \sqrt 3 a$ 
    inspired by the nesting vector obtained from the generalized 
    susceptibility.
    }
    \label{fig:OOGGA}
\end{figure}
%%%%%%%FIG1
With the new supercell, we allow for a small
displacement of Fe atoms (according to Ref.~\onlinecite{Gironcoli})
to check for the possibility of relaxation towards a new minimum
in the GGA approximation. This new state is orbital ordered 
insulating state, as can be seen in the charge density profile shown
in Fig.~\ref{fig:OOGGA}. The insulating behavior of this state
is due to broken symmetry accompanying the orbital ordering. 
The gap magnitude at $\Gamma$ point for this state is $E_g^{\rm GGA}\simeq0.54$ eV. 
The energy of this new minimum per supercell is about $0.47$ eV 
lower than the corresponding non-ordered metallic state.

  Although in the GGA approximation when we allow for the possibility
of a broken symmetry state, we find an insulating state with lower energy
than the original metallic one, but the energy gap is still
far less than the experimentally observed value of $E_g=2.4$ eV~\cite{Bowen}.
The GGA energy level of FeO suggests the following picture: Five 
majority spins are well below the Fermi level, leaving one minority
spin in the relevant $t_{2g}$ band~\cite{Gironcoli}. To this extent, we consider
FeO to belong to $d^1$ class. 
Therefore we are dealing with one
minority spin per site Mott insulating situation. Hence, we have
to take into account the role of on-site Coulomb interaction, $U$.
We incorporate the effect of $U\sim 4.3$ eV
within the GGA+U approximation. With this value of $U$,
we obtain the energy gap $E_g^{\rm GGA+U}\sim 2.2$ eV at the $\Gamma$ point.
Note that we did not adjust the value of $U$. The above value for $U$ was
calculated in Ref.~\onlinecite{Gironcoli}.
Compared to sophisticated quantum Monte Carlo calculations~\cite{MitasFeO}, 
our result is in remarkable agreement with the experimental value~\cite{Bowen}.
Therefore both orbital ordering, as well as Coulomb correlations
are important to understand the electronic structure of FeO
Mott insulator.

%%%%%%%FIG1
\begin{figure}[t]
    \includegraphics[width=8cm,angle=0]{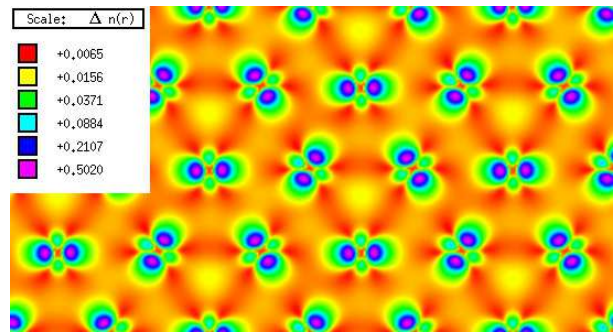}
    \caption{(Color online) Orbital ordering in the GGA+U solution.
    This pattern is $d^1$ analogue of the $d^2$ mean field state
    proposed in Ref.~\onlinecite{Pen} for LiVO$_2$. As can be seen
    the incorporation of correlation effects via $U$, does 
    not destroy the orbital ordering.
    This figure corresponds to the state with
    $(\alpha,\beta,\gamma)=(1,0,0)$.
    }
    \label{fig:OOGGAU}
\end{figure}
%%%%%%%FIG1

%table1!!!!!!!!!!!!!!!!!!!!!!!!!!!!!!!!!!!!!!!!!!!!!!!!!!!!!!!!!!!!!!!!!!!!!!!!!!!!!!!!!!!!!!!!!!!!!!!!!!!
\begin{table}[b]
\caption{Energy gained by orbital ordering with respect to non-ordered $a_{1g}$
state. For  each case we include the energy of system without atomic position relaxation
and with atomic position relaxation. The energies are expressed as energy per unit cell 
(each cell contains six Iron atoms).
	 }
\label{tab:OO}
\begin{ruledtabular}
\begin{tabular}{c|c|c}
Configuration                      &  $\rm E- \rm E_{a_{1g}}$(meV)       & $\rm E-\rm E_{a_{1g}}$(meV) \\
$(\alpha,\beta,\gamma)$            &   without relaxation            &   with  relaxation \\ \hline
$(1,0,0)$                          &           $-60$                &      $-174$       \\ \hline
$\frac{1}{\sqrt{2}}(1,1,0)$        &           $-36$                &     $ -66$        \\  
\end{tabular}
\end{ruledtabular}
\end{table}
%!!!!!!!!!!!!!!!!!!!!!!!!!!!!!!!!!!!!!!!!!!!!!!!!!!!!!!!!!!!!!!!!!!!!!!!!!!!!!!!!!!!!!!!!!!!!!!!!!!!!!!!!!!
%table1!!!!!!!!!!!!!!!!!!!!!!!!!!!!!!!!!!!!!!!!!!!!!!!!!!!!!!!!!!!!!!!!!!!!!!!!!!!!!!!!!!!!!!!!!!!!!!!!!!!

With GGA+U there is a freedom on choosing the appropriate state in the $t_{2g}$ 
subspace to be occupied by minority spin.
The most general state in this subspace is of the from, 
$\alpha\vert xy \rangle+\beta\vert yz \rangle+ \gamma\vert zx \rangle$.
The underlying triangular lattice structure with $C_3$ symmetry requires
invariance under $120^\circ$ rotations. Therefore the two other states
are required to be,
$\alpha\vert yz \rangle+\beta\vert zx \rangle+ \gamma\vert xy \rangle$,
$\alpha\vert zx \rangle+\beta\vert xy \rangle+ \gamma\vert yz \rangle$.
Each set of values for $(\alpha,\beta,\gamma)$ corresponds to a new
orbital ordered configuration. The only choice which leads to state
without orbital ordering is the homogeneous one with $|\alpha|=|\beta|=|\gamma|=1/\sqrt 3$ 
(the so called $a_{1g}$ state).
In GGA+U it is not feasible to check the energy of all possible states.
In table~\ref{tab:OO} we compare the energy of 
the non-ordered $a_{1g}$ state with two states  corresponding to 
$(\alpha,\beta,\gamma)=(1,0,0)$, and
$(\alpha,\beta,\gamma)=(1,1,0)/\sqrt 2$.
Second column of table~\ref{tab:OO} indicates that, in presence of
electron-electron interaction $U$, orbital ordering still reduces the total
energy by tens of meV per unit cell. The state $(1,0,0)$ has lower
energy for which we have plotted the charge density in Fig.~\ref{fig:OOGGAU}.
This state is the negation of the state proposed in Ref.~\cite{Pen}
for the LiVO$_2$. Such a negation image is reasonable given the fact that
$d^1$ and $d^2$ classes in $t_{2g}$ subspace are connected with 
a particle-hole symmetry transformation.

%It is well known that the orbital ordering is always accompanied by
%appropriate Jahn-Teller distortion. 
So far we have checked that electronic interactions can take
advantage of the instability suggested by the nested Fermi surface,
to stabilize an orbital ordered pattern. Next we ask the question,
can electron-lattice interactions take advantage of this instability
to give rise to appropriate form of Jahn-Teller distortion?
To verify this for the case of FeO, we start with  orbital ordered
states $(\alpha,\beta,\gamma)$, then we allow the atomic positions
to relax.
In the third column of table~\ref{tab:OO}, we report the calculated
total energy per unit cell when the relaxation is allowed.
As it can be seen, the relaxed structure has
lower energy than the corresponding non-relaxed state.
Note that to verify whether the orbital ordered state is 
more stable than the non-ordered state, we first distorted
the atomic positions which resulted in orbital ordering pattern
of Fig.~\ref{fig:OOGGA}. But here we check the reverse sequence, i.e.
we start off with an orbital ordered state, then we check whether the
atomic displacements can further stabilize orbital ordered state or not.
Therefore for both orbital ordered states
considered here, the ordering is always accompanied with cooperative 
Jahn-Teller distortion.

To summarize, we started with a GGA metallic band picture.
The inter-band nesting pointed us to examine the energy of 
orbitally ordered state within the same approximation. 
However,
since in the case of transition metal oxides the Coulomb correlation
U is also important, we also compared the energy of orbital ordered
states in presence of on-site Coulomb interaction U, in the GGA+U
approximation. In FeO, Orbital ordering significantly stabilizes both
GGA as well as GGA+U states with respect to corresponding non-ordered
phases. We also allowed for relaxation of the atomic positions on 
top of orbitally ordered GGA+U state. We found that the Jahn-Teller distortion
accompanying the orbital ordering is automatically realized in our 
approach. As a check of this method, we also studied the cases
of LiVO$_2$ and LaMnO$_3$ which are known to exhibit orbital
ordering at low temperatures. Therefore we propose this Fermi surface analysis
as a rout to explore the possibility of orbital ordering and/or 
reduction in the lattice symmetry, within the DFT (LDA/GGA) electronic structure calculations.

{\em Acknowledgement}: The authors are grateful to S. Maekawa 
for critical proof reading of the paper. M.A. wishes to thank
H. Akbarzadeh.


\begin{thebibliography}{50}
\bibitem{tech-details} 
To calculate the generalized susceptibility we used $120 \times 120 \times 120$ 
k-point mesh for FeO. For $\rm LiVO_{2}$ and $\rm LaMnO_{3}$ we used
$100 \times 100 \times 100$ and $100 \times 100 \times 50$ 
mesh respectively. Since the $\rm LiVO_{2}$ is spin singlet, 
we assume a nonmagnetic solution, while in FeO with the same
lattice structure (trigonal) we assume magnetic solution.
For $\rm LaMnO_{3}$ we use a 
tetragonal structure for the A-type phase.
\end{thebibliography}
\end{document}